\documentclass[copyright,creativecommons]{eptcs}

\usepackage{amsmath,amsfonts,amssymb,graphics,graphicx,epsfig,color,times,dsfont}

\providecommand{\event}{QPL 2011} % Name of the event you are submitting to
\usepackage{breakurl}             % Not needed if you use pdflatex only.

\title{Generalizations of Boxworld}
\author{Peter Janotta
\institute{Fakult{\"{a}}t f\"{u}r Physik und Astronomie\\Universit\"{a}t W\"{u}rzburg\\Am Hubland\\97074 W\"{u}rzburg\\Germany}
\email{peter.janotta@physik.uni-wuerzburg.de}
}

\def\titlerunning{Generalizations of Boxworld}
\def\authorrunning{P. Janotta}

\begin{document}
\maketitle

\begin{abstract}
Boxworld is a toy theory that can generate extremal nonlocal correlations known as PR boxes. These have been well established as an important tool to examine general nonlocal correlations, even beyond the correlations that are possible in quantum theory. We modify boxworld to include new features. The first modification affects the construction of joint systems such that the new theory allows entangled measurements as well as entangled states in contrast to the standard version of boxworld. The extension to multipartite systems and the consequences for entanglement swapping are analysed. Another modification provides continuous transitions between classical probability theory and boxworld, including the algebraic expression for the maximal CHSH violation as a function of the transition parameters. 
\end{abstract}

\section{Introduction}

It is well known that quantum theory allows so-called nonlocal correlations, i.e. correlations that cannot be produced by any local hidden variable theory \cite{Bell, chsh}.
However, Popescu and Rohrlich constructed non-signalling correlations even beyond those possible in quantum theory, called PR boxes \cite{PR}.
Various powerful features have been revealed for PR boxes. 
For example the PR boxes provide a nonlocal resource that would make communication complexity trivial \cite{vanDam}, they can violate information causality \cite{IC} and exceed the power of quantum resources for some nonlocal games \cite{chsh, noah, GYNI}, just to name a few.

There have been various attempts to generalize entanglement swapping, known from quantum mechanics \cite{qmswap}, to a swapping protocol for PR boxes, i.e. a LOCC protocol that generates a PR box between two parties Alice and Charlie with the help of a third party (Bob) who shared an PR box with Alice and Charlie individually.
A popular approach is known as \emph{nonlocality swapping}.
In contrast to entanglement swapping in quantum mechanics nonlocality swapping does not transform entangled states directly but introduces a so-called coupler that acts on probability distributions on Bob's side.
It was shown in \cite{NoPRswap} that being able to apply the same coupler consistently to all PR boxes does not allow any nontrivial coupler for two-input-two-output boxes, i.e. nonlocality swapping is impossible for that case.
By exclusion of all PR boxes but one, however, it was shown in \cite{genbox} that nonlocality swapping can be done. 
The other PR boxes can be generated by a simple relabelling of measurements and outcomes such that they can be generated from the single box by mere classical local wirings.
That the coupler is not sensitive to wirings is explained by introducing a genuine part of the box that corresponds to the (not further specified) state that gives rise to the correlations represented by a nonlocal box.
The details on the underlying entanglement swapping on physical states that produces PR box correlations remain unclear.

It is known that PR box correlations can be generated by a toy theory referred to as \emph{boxworld} \cite{PRtrans}.
In \cite{swapboxworld} it was shown that entanglement swapping in boxworld is not possible.
This results from the fact that the standard version of boxworld does not allow any entangled measurements.
We show in section \ref{sec:modone} that one can modify joint systems in boxworld such that entangled states as well as entangled measurements are included.
We study whether this construction can be extended to multipartite systems and show that this requires a different treatment of bipartite partitions consisting of otherwise equal elementary systems.
It is demonstrated that this modified theory shows potential for entanglement swapping.

In \cite{Beigi09,locallimits} it was shown that the local structure of a theory can limit which bipartite correlations are possible. 
This is further studied in a second modification of boxworld in section \ref{sec:modtwo}.
We provide a continuous transition between classical probability theory and the standard version of boxworld in section \ref{sec:modtwo} by modifying the local structure of the theory.
This could be used as a tool to study the emergence of nonlocal correlations.
The analytic expression for the maximal CHSH-violation as a function of the transition parameters is given.

\section{The no-signalling polytope}

When considering correlations we are interested in measurement statistics.
These are represented by probability distributions $p(ab|xy)$ to get outcome $a$ and $b$ given that local measurements $x$ and $y$ have been chosen.
The correlations are supposed to occur independently of the separation of the two parts. 
In particular the parts could be space-like separated, i.e. communication between the parts is regarded impossible. 
Excluding correlations that would allow superlumial transmission of information, requires the marginal probability distributions to be independent of the choice of the local measurement on the other side of the system.
Formally, this refers to the \emph{no-signalling principle}
\begin{align}
 p(a|x) &= \sum_b p(ab|xy) \quad \forall y\\
 p(b|y) &= \sum_a p(ab|xy) \quad \forall x.
\end{align} 

The simplest set-up where nonlocal correlations occur is for two choices of binary local measurements at each side of a bipartite system. 
For this situation the set of all correlations that can be constructed by a local hidden variable theory is given by probability distributions satisfying the CHSH inequality \cite{chsh}
\begin{align}
 \label{eq:CHSHinequalty}
 |C_{00}&+C_{01}+C_{10}-C_{11}| \leq 2\\
 \nonumber C_{xy} &= \sum_{a,b} (-1)^{a \oplus b} \, p(ab|xy), 
\end{align}
and equivalent inequalities where the measurements and outcomes have been relabelled.

Together with trivial conditions $p(ab|xy) \in [0,1]$ on the range of probabilities this defines the so-called \emph{local polytope} $\mathcal{L}$ spanned by extremal correlations $p(ab|xy)=p(a|x) \, p(b|y)$ with $p(a|x), p(b|y) \in \{0,1\}$.
Restricting correlations only by the no-signalling principle, however, allows additional extremal correlations, the so-called PR-boxes of the form
\begin{align}
\label{eq:PRbox}
p(a,b|x,y) =
\begin{cases}
\frac{1}{2} & \text{if $ a \oplus b = xy$} \\
0 & \text{otherwise}
\end{cases}
\end{align}
and equivalent probability distributions with relabelled measurements and outcomes.
These boxes give the algebraic maximum of $4$ for the term in \eqref{eq:CHSHinequalty}, i.e. violates the CHSH inequality.
The PR boxes together with the extremals of the local polytope span the no-signalling polytope $\mathcal{P}$ that includes all correlations that respect the no-signalling principle.

The set of quantum correlations $\mathcal{Q}$ lies between the local and the no-signalling polytope and has infinitely many extremal points, i.e. is not a polytope.
The maximum CHSH violation possible for quantum resources is given by Tsirelson's bound of $2 \sqrt{2}$ \cite{cirelson80}.
Although various approximations to $\mathcal{Q}$ have been derived \cite{NPA, momentproblem, Uff02} the complete structure of the set of quantum correlations is still unknown.
Notably, there are correlations in $\mathcal{P}$ not possible to produce within quantum theory.
These post-quantum correlations are regarded to be unphysical.
Nevertheless, they have been a subject of extensive research, since they help to understand nonlocal correlations in general.
It is an open question whether there is a single physical principle that explains the absence of post-quantum correlations, but at the same time allows the nonlocal correlations that can be explained within quantum theory.

\section{Generalized probabilistic theories}
\label{sec:GPTs}

Usually post-quantum correlations are studied on the basis of probability distributions, without any reference to physical theories that allow to produce them.
Beyond that, the framework of generalized probabilistic theories (GPTs) can establish interesting connections between toy theories and the correlations allowed therein \cite{Beigi09,locallimits,Acin10}.
This section introduces the basic notions and concepts of the framework.
It can be regarded as a generalization of the density matrix formalism in quantum theory to arbitrary state space geometries.

\subsection{Single systems}
\label{sec:GPTsingle}

States $\omega$ are objects that determine the measurement statistics for any possible measurement on a physical system.
The framework is operational, i.e. situations with the same measurement statistics on a physical system for any measurement refer to the same state. 
Similar to density matrices in quantum theory, preparing different states $\omega_i$ with probability $p_i$ results in the mixed state $\omega = \sum_i p_i \, \omega_i$ given by a convex combination of the states $\omega_i$.
The set of all states allowed in a theory is the convex state space $\Omega$.
\emph{Pure states} are extremal points of $\Omega$, i.e. points that cannot be written as a convex combination of other states.

A measurement outcome $i$ of measurement $M$ corresponds to a so-called effect $e_i^M$. 
This is a linear functional $e_i^M: \Omega \to [0,1]$ that map a state $\omega$ to the probability $p(i|\omega, M) = e_i^M (\omega)$ to get the related measurement outcome.
The unit measure $u$ is the unique effect $u: \Omega \to 1$ that corresponds to a measurement with only one outcome that is certain to occur for every state.
Arbitrary measurements are given by set of effects that sum up to the unit measure, i.e. produce a normalized probability distribution for any state.
Particular simple measurements are given by binary measurements that are generated by a single effect $e$ and its complement $\bar{e} = u - e$.

In order to identify a state uniquely one does not need the measurement statistics for every possible measurement, but only for a finite number of binary measurements, the so-called fiducial measurements.
The minimal number of binary measurements needed to identify any state uniquely is the dimension of the linear spaces spanned by effects.
In this work we restrict ourselves to the finite-dimensional case.

It is sometimes useful to cope with unnormalized states $\tilde{\omega} = \lambda \, \omega$, i.e. a state scaled by a non-negative factor $\lambda \geq 0$.
The set of unnormalized states form a \emph{positive cone} $A_+$. 
Since, applying effects to states should result in a positive value, the set of effects must be a subset of the \emph{dual cone}
\begin{align}
 A^*_+ := \{ e: e(\tilde{\omega}) \geq 0 \, \forall \tilde{\omega} \in A_+ \}
\end{align} 
Systems with a one-to-one correspondence between elements of the positive cone and the dual cone are called \emph{weakly self-dual}.
Such systems are of special interest, since weak self-duality is necessary for important information theoretic tasks like steering or teleportation \cite{steering, teleport}.

\subsection{Joint systems}

Besides the description of single systems a theory has also to define composed systems, i.e. which joint states are allowed.
There should at least be situations included where the preparation of local states and the performing of local measurements are done independently in each subsystem.  
This causes the state space $\Omega^{AB}$ of joint system $AB$ to be chosen between two extremes, the \emph{minimal tensor product} and the \emph{maximal tensor product}.
Normalization of joint states is given with respect to the joint unit measure $u^{AB} = u^A \otimes u^B$.
A general joint state $\omega^{AB}$ is given by an affine combination of product states, i.e.
\begin{align}
 \omega^{AB} = \sum_{i} \lambda_i \, \omega^A_i \otimes \omega^B_i \text{ with } \sum_i \lambda_i = 1.
\end{align}
If all the coefficients $\lambda_i$ are positive the affine combination reduces to a convex combination and the state $\omega^{AB}$ is separable.

The \emph{minimal tensor product} describes a joint system that contains only separable states, i.e. all product states $\omega^A \otimes \omega^B$ with $\omega^{A/B} \in \Omega^{A/B}$ and the convex combinations of these states.
This represents the situation where the preparation of each subsystem is independent or classically correlated. 

Similar to the independent preparation of states it should also always be possible to perform measurements independently to each subsystem.
Thus, a theory has to include product effects $e^A \otimes e^B$.
Since applying such an effect to any joint state should result in a probability, the set of joint states is restricted to give positive values.
This includes all states of the minimal tensor product as well as additional states.
The set of all states that satisfy this condition is called the \emph{maximal tensor product}.
Joint states that are not included in the minimal tensor product are called \emph{entangled}. 

The tensor product defines the positive cone of the joint system.
Joint effects are elements of the dual cone.
For finite dimensions, the maximal tensor product for the states leads to the minimal tensor product for the dual cone, and the other way round, the minimal tensor product for states leads to the maximal tensor product for the effects.

The specification of a tensor product that can be chosen between the minimal and the maximal tensor product is part of the definition of a theory.

\subsection{Examples}

\subsubsection{Classical probability theory}
In classical probability theory every pure state $\omega_i$ gives rise to an extremal effect $e_i$ that identifies it uniquely, i.e. $e_i (\omega_j) = \delta_{ij}$.
Therefore, the dimension of the system is equal to the number of pure states, corresponding to linear independent vectors that form a basis. 
The representation of an arbitrary (mixed) state with respect to that basis is its unique decomposition into pure states.
The state spaces are given by simplices.

For the joint systems the minimal and the maximal tensor product coincide.
Hence, there are neither entangled states nor measurements.

\subsubsection{Quantum theory}

In quantum theory both states and effects are given by positive hermitian matrices.
Applying an effect to a state is done by the Hilbert-Schmidt inner product.
The unit measure is the identity matrix.
Thus, normalized states $\rho$ have trace one.

Joint states and effects are again given by positive Hermitian operators (just of higher dimension). 
The condition that positivity is preserved in joint systems defines the tensor product.
It strictly lies between the minimal and the maximal tensor product.
The tensor product allows entangled states as well as entangled effects and there is a one-to-one mapping between them.

\subsubsection{Boxworld}
  
As introduced in \ref{sec:GPTsingle} the state of a system can be identified by a set of fiducial measurements.
In boxworld state spaces of single systems include states with measurement statistics in the full range $[0,1]$ for every fiducial measurement independently. 
The extremal points are deterministic states that give a certain outcome $p_e, p_{\bar{e}} \in \{0,1\}$ for any fiducial measurement, i.e these pure states show no uncertainty for the fiducial measurements.
The state space is a hypercube. 

For the CHSH set-up with two parties each choosing between two binary measurements the case with only two fiducial measurements is of special interest.
In this case the state space is given by a two-dimensional square.
A possible representation of the extremal points is given by the following vectors in $\mathds{R}^3$:
\begin{align}
 \omega_1 = \begin{pmatrix} 1 \\ 0 \\ 1 \end{pmatrix} \quad
 \omega_2 = \begin{pmatrix} 0 \\ 1 \\ 1 \end{pmatrix} \quad
 \omega_3 = \begin{pmatrix} -1 \\ 0 \\ 1 \end{pmatrix} \quad
 \omega_4 = \begin{pmatrix} 0 \\ -1 \\ 1 \end{pmatrix}
\end{align}
The set of effects is given by the zero-functional, the unit measure $u$ and the following extremal effects:
\begin{align}
 u = \begin{pmatrix} 0 \\ 0 \\ 1 \end{pmatrix} \quad
 e_1 = \frac{1}{2} \begin{pmatrix} -1 \\ -1 \\ 1 \end{pmatrix} \quad
 e_2 = \frac{1}{2} \begin{pmatrix} 1 \\ -1 \\ 1 \end{pmatrix} \quad
 e_3 = \frac{1}{2} \begin{pmatrix} 1 \\ 1 \\ 1 \end{pmatrix} \quad
 e_4 = \frac{1}{2} \begin{pmatrix} -1 \\ 1 \\ 1 \end{pmatrix}
\end{align}
Note that there is a one-to-one correspondence between extremal states $\omega_i$ and extremal effects $e_i$. 
Namely, rotating any $e_i$ by $\frac{3 \pi}{4}$ round the z-axis and proper scaling yields the corresponding pure state $\omega_i$.
I.e. the single system introduced above is weakly self-dual.

For the joint system one allows all states compatible with the local measurements, i.e. the maximal tensor product.
As a consequence the set of joint effects is limited to separable ones. 
The joint state space $\Omega^{AB}_{max}$ is spanned by the $16$ extremal product states and $8$ additional entangled pure states:
\begin{align}
 \omega^{AB}_{17}& = \frac{1}{2} \left( \omega^A_1 \otimes \omega^B_2 - \omega^A_2 \otimes \omega^B_2 + \omega^A_2 \otimes \omega^B_3 + \omega^A_3 \otimes \omega^B_1 \right)\\
 \omega^{AB}_{18} &= \frac{1}{2} \left( \omega^A_2 \otimes \omega^B_2 - \omega^A_3 \otimes \omega^B_3 + \omega^A_3 \otimes \omega^B_4 + \omega^A_4 \otimes \omega^B_3 \right)\\
 \omega^{AB}_{19} &= \frac{1}{2} \left( \omega^A_1 \otimes \omega^B_1 - \omega^A_2 \otimes \omega^B_2 + \omega^A_2 \otimes \omega^B_3 + \omega^A_3 \otimes \omega^B_2 \right)\\
 \omega^{AB}_{20} &= \frac{1}{2} \left( \omega^A_2 \otimes \omega^B_2 - \omega^A_3 \otimes \omega^B_2 + \omega^A_3 \otimes \omega^B_3 + \omega^A_4 \otimes \omega^B_1 \right)\\
 \omega^{AB}_{21} &= \frac{1}{2} \left( \omega^A_1 \otimes \omega^B_4 - \omega^A_1 \otimes \omega^B_1 + \omega^A_2 \otimes \omega^B_1 + \omega^A_4 \otimes \omega^B_2 \right)\\
 \omega^{AB}_{22} &= \frac{1}{2} \left( \omega^A_1 \otimes \omega^B_4 - \omega^A_1 \otimes \omega^B_1 + \omega^A_2 \otimes \omega^B_2 + \omega^A_4 \otimes \omega^B_1 \right)\\
 \omega^{AB}_{23} &= \frac{1}{2} \left( \omega^A_1 \otimes \omega^B_1 - \omega^A_1 \otimes \omega^B_2 + \omega^A_2 \otimes \omega^B_2 + \omega^A_4 \otimes \omega^B_3 \right)\\
 \omega^{AB}_{24} &= \frac{1}{2} \left( \omega^A_1 \otimes \omega^B_1 - \omega^A_1 \otimes \omega^B_4 + \omega^A_2 \otimes \omega^B_4 + \omega^A_4 \otimes \omega^B_3 \right)
\end{align}

It is well known that this theory is able to generate correlations spanning the whole no-signalling polytope.
In particular there are measurements on the entangled pure states that generate the PR boxes \eqref{eq:PRbox}. 
Notably, this is also possible with each of the nonlocal extremal states individually.
Any no-signalling correlations can be generated by measurements on a single pure entangled state.

In the next sections we will modify this standard construction of boxworld.
We will construct another tensor product that has not only entangled states, but entangled measurements as well.
Notably, this will be done without loosing any of the correlations possible.
On the other hand by keeping all joint states possible, but changing the single system we will provide a transition between classical probability theory with no nonlocal correlations at all and standard boxworld allowing any non-signalling correlations.

\section{A weakly self-dual version of boxworld}
\label{sec:modone}

Systems with a one-to-one correspondence between the positive cone of unnormalized states and the dual cone, are referred to as \emph{weakly self-dual} systems.

In boxworld single systems are weakly self-dual.
However, for joint systems, in the standard formulation of boxworld with the maximum tensor product, the systems lose this property.
Joint systems include all possible joint states, but no entangled effects.
We would like to study a situation equivalent to quantum theory where entanglement in states and measurements is balanced with a one-to-one correspondence between effects and unnormalized states.

In order to find such a weakly self-dual tensor product we take a closer look on the tensor product of two qubits and try to mimic it.
Exchanging the usual tensor product of two qubits by the maximal tensor product, one gets a joint state space that still includes all pure states available in standard quantum theory. 
However, it inhabits also some additional pure states that each can be generated by a partial transpose applied to a corresponding entangled pure quantum state. 
It is well known that such states are non-positive and therefore normally excluded \cite{Horodecki}.
I.e. each entangled state in a standard bipartite qubit system has a counterpart in the maximal tensor product that is regarded unphysical.
At the same time each new state leads to new constraints on the entangled effects possible.
From this point of view the standard tensor product of two qubits can be derived from the maximal tensor product by excluding half of the entangled pure states.
This removes constraints on corresponding entangled effects.
  
We construct a weakly self-dual tensor product for local boxworld systems in a similar manner. 
We include separable pure states and only four ($\omega^{AB}_{17}, \omega^{AB}_{18}, \omega^{AB}_{19}, \omega^{AB}_{20}$) of the nonlocal extremals as pure joint states.
The reduction of the extremal entangled states, allows for new extremal entangled effects, such that joint states and effects are connected by a one-to-one mapping:
\begin{align}
\label{eq:enteffects}
 e^{AB}_{17}& = \frac{2}{3} \left( e^A_1 \otimes e^B_2 - e^A_2 \otimes e^B_2 + e^A_2 \otimes e^b_3 + e^A_3 \otimes e^B_1 \right)\\
 e^{AB}_{18} &= \left(e^A_2 \otimes e^B_2 - e^A_3 \otimes e^B_3 + e^A_3 \otimes e^B_4 + e^A_4 \otimes e^B_3 \right)\\
 e^{AB}_{19} &= \left(e^A_1 \otimes e^B_1 - e^A_2 \otimes e^B_2 + e^A_2 \otimes e^B_3 + e^A_3 \otimes e^B_2 \right)\\
 e^{AB}_{20} &= \frac{2}{3} \left( e^A_2 \otimes e^B_2 - e^A_3 \otimes e^B_2 + e^A_3 \otimes e^B_3 + e^A_4 \otimes e^B_1 \right)
\end{align}
Nevertheless, since we keep some of the original entangled pure states, we are still able to produce PR boxes with all the powerful features that come with them.

Dynamics are connected to automorphisms of the state space. 
Gross et al. have shown that dynamics in standard boxworld are very limited, since invertible maps of the joint systems include only permutations of the local pure states \cite{PRtrans}.
This means that there is no reversible way to convert pure separable states to pure entangled states, e.g. as it is possible in quantum theory with the CNOT operation.
If one want to include at least one of the original entangled pure states of boxworld, one can get the others by the local symmetries.
Thus, excluding some of the entangled extremals leads to a situation where these local symmetries would map joint states allowed in the theory to those that have been excluded.
This gives rise to a notion of complete positivity as opposed to positivity, similar to the situation in quantum theory.
As a consequence the construction above restricts dynamics even further.

\subsection{Multipartite extension and entanglement swapping}

Given the new tensor product, we want to extend it to multipartite systems as well. 
Using the example of a four-partite system $ABCD$, we will show in this section that this, if possible at all, can only be done under conditions that could be regarded unphysical, namely the symmetry under swapping of the identical elementary systems has to be broken.

No matter what the exact structure of the four-partite tensor product might be, it must at least contain any product state consisting of two bipartite states in $AB$ and $CD$ of the bipartite tensor product chosen before.
Assuming that the systems $BD$ allow for the same joint measurements, it can be shown that applying the entangled effects in \eqref{eq:enteffects} to the product $\omega^{AB} \otimes \omega^{CD}$ of two nonlocal extremal states would result in a collapse to an entangled pure state in $AC$, i.e. the realization of entanglement swapping.
The resulting state $\omega^{AC}$ in $AC$ however is not guaranteed to be in the bipartite tensor product chosen before.
One can find combinations of states and effects such that $\omega^{AC}$ is one of the states of the maximal tensor product that we abolished for the construction of our new tensor product.
For example if one gets outcome $e^{BD}_{17}$ when measuring the state $\omega^{AB}_{17} \otimes \omega^{CD}_{18}$ the resulting collapsed state in $AC$ is $\omega^{AC}_{22}$ that is not included in the weakly self-dual tensor product chosen before.
This means that the bipartite tensor product for subsystem $BD$ and $AC$ must be different to the tensor product chosen for $AB$ and $CD$.
Hence, the initial construction can only be extended in a way that does not treat each pairs of particles in the same way.

Since elementary systems of the same form should be indistinguishable, it is questionable if this is consistent with a reasonable notion of a multipartite system.
More formally, this shows that the construction of the weakly self-dual tensor product above cannot be extended to an associative tensor product.
Note however, that the construction of a weakly self-dual tensor product is not unique.
Therefore, it might still be possible to find another weakly self-dual tensor product that is associative, can generate PR boxes and allows entanglement swapping of the included nonlocal extremal states at the same time.

\section{Transition from classical probability theory to boxworld}
\label{sec:modtwo}

In order to provide a tool to examine the emergence of nonlocal correlations, this section introduces another modification of boxworld.
It provides a transition from classical probability theory to the original boxworld case. 
This is done by exchanging one of the original extremal states $\omega_1$ with
\begin{align}
 \omega'_1 = \begin{pmatrix} x \\ y \\ 1 \end{pmatrix} \qquad x \in [0,1], y \in [x-1, 1-x],
\end{align}
that can be varied conditioned on the two transition parameters $x$ and $y$.
For $x = 0$ one gets the classical case, whereas $x = 1, y = 0$ corresponds to standard boxworld.

\begin{figure}[h]
\centering
\includegraphics[width=.25 \linewidth]{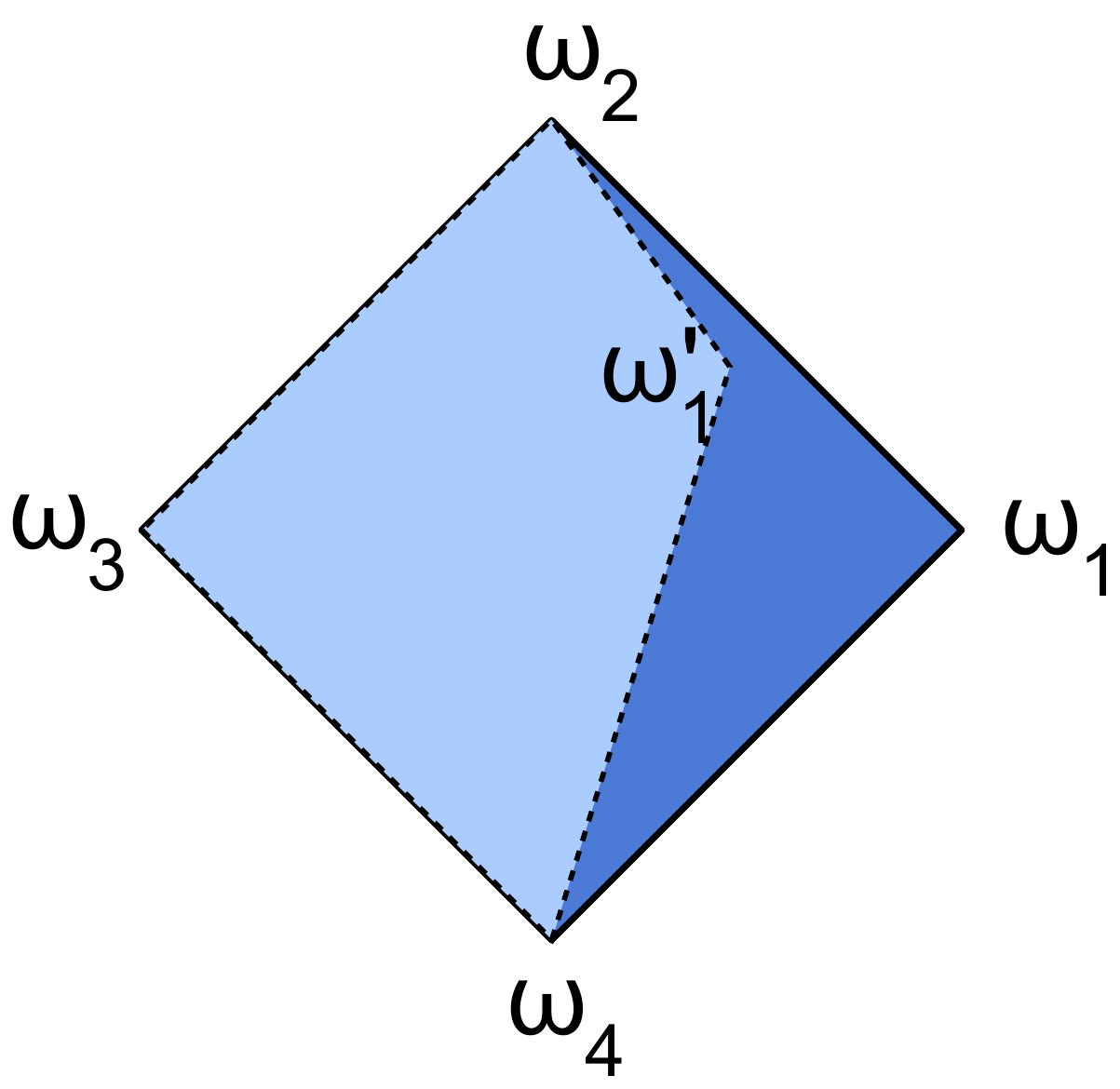}
\caption{State space of single systems in modified boxworld}
\end{figure}

Looking at bipartite systems consisting of two of these single systems, it possible to calculate the extremal joint states (again $16$ separable and $8$ entangled) of the maximal tensor product as a function of $x, y$.
Since the number of different extremal measurements and extremal joint states is rather small, the maximal CHSH-value of the theory can be found by brute-force.

It is given by:
\begin{align}
2+\frac{16 x^2}{x^2+(|y|-1)^2+2 x (3+|y|)}
\end{align}

\begin{figure}
\centering
\includegraphics[width=.65 \linewidth]{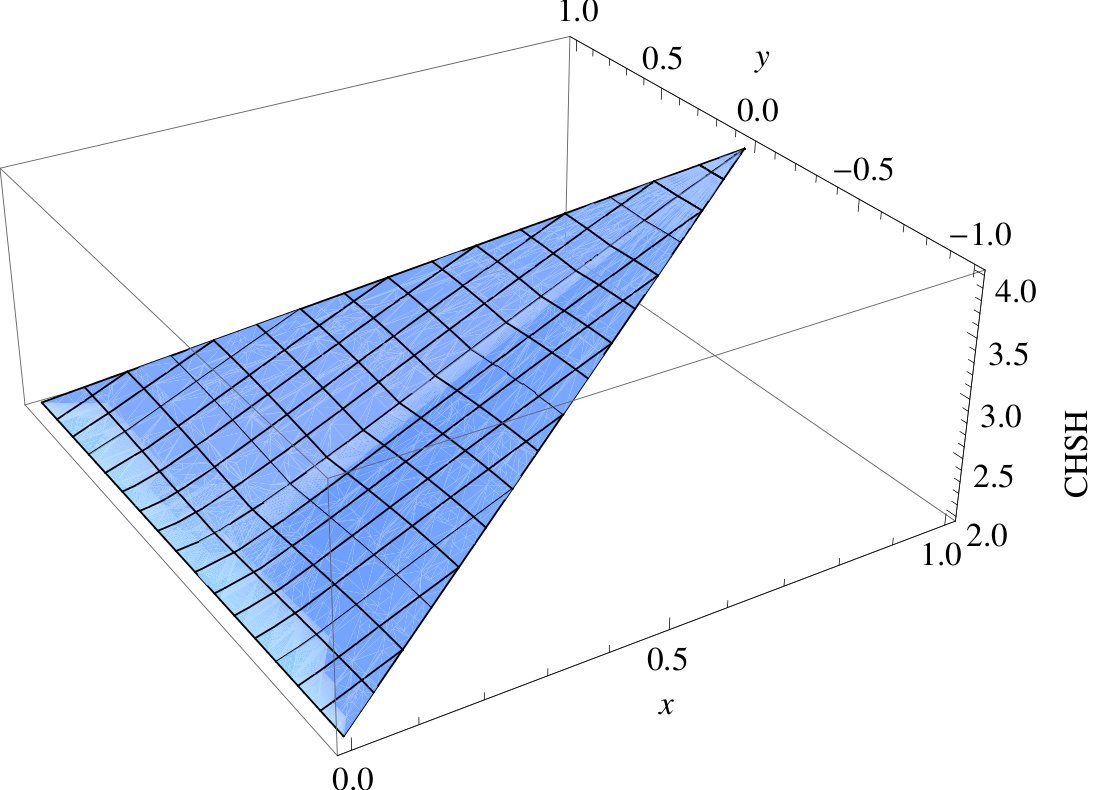}
\caption{Maximal CHSH-violation as a function of the transition parameters $x$ and $y$}
\end{figure}

One can also show that these systems allow correlations not even possible in quantum theory for arbitrary small derivations of the classical state space.
This complements a result by Brunner et. al. \cite{BS} that there are correlations arbitrary close to classical correlations that are not possible in quantum theory.
This work shows that such correlations can be gained by theories with local state spaces including only states that are arbitrary close to those allowed in classical probability theory.

\section{Future work}

The results presented here are work-in-progress. 
For the first modification it is still open if the extension of the bipartite weakly self-dual tensor product to multipartite systems can be weakly self-dual again, even if we accept the broken swapping symmetry shown in section \ref{sec:modone}.
Also, it has to be addressed if there are any other inequivalent choices for an weakly self-dual tensor product that allow PR boxes and if these might be associative.
Concerning the second modification, it would be interesting to classify the exact set of possible correlations for the systems introduced in section \ref{sec:modtwo} as a function of the transition parameters, not only the maximal CHSH violations.
Future work will also try to unify the two modifications of boxworld presented here.

\section*{Acknowledgements}
The author is grateful for discussions with Christian Gogolin, Volkher Scholz, Howard Barnum and Jonathan Barrett.

\bibliography{refs}
\bibliographystyle{eptcs}

\end{document}